\documentclass[fleqn,11pt]{article}
\usepackage{amsmath,amssymb}
\usepackage{xcolor}
\usepackage{verbatim}

\def\stackunder#1#2{\mathrel{\mathop{#2}\limits_{#1}}}

\newcommand{\Lee}[1]{\stackunder{#1}{\mathfrak{L}}}

\def\stackunder#1#2{\mathrel{\mathop{#2}\limits_{#1}}}
\newcommand{\Fig}[3]{%
\begin{center}
\parbox{8cm}{%
\refstepcounter{figure}\includegraphics[width=8cm,height=#2cm]{#1} \noindent Fig. \thefigure:\quad
#3}\end{center}}

\newcounter{strochka}

\newcounter{spisok}
\usepackage{amsfonts,amssymb,cite}
\usepackage{graphicx}

\def\noi{\noindent}

\newcommand{\Title}[1]{\noi {{\Large\bf #1}}\\[1ex]}

\newcommand{\Author}[2]{\noi{\bf #1}\\[2ex]\noi{\normalsize\it #2}\\}

\newcommand{\Abstract}[1]{\vskip 2mm \begin{center}
        \parbox{16.4cm}{\small\noi #1} \end{center}\medskip}

\def\nqq{\hspace*{-2em}}

\usepackage{color}

\def\Jl#1#2{#1 {\bf #2},\ }

\def\ApJ#1 {\Jl{Astroph. J.}{#1}}
\def\CQG#1 {\Jl{Class. Quantum Grav.}{#1}}
\def\DAN#1 {\Jl{Dokl. AN SSSR}{#1}}
\def\GC#1 {\Jl{Grav. Cosmol.}{#1}}
\def\GRG#1 {\Jl{Gen. Rel. Grav.}{#1}}
\def\IJMPD#1 {\Jl{Int. J. Mod. Phys. D}{#1}}
\def\JETF#1 {\Jl{Zh. Eksp. Teor. Fiz.}{#1}}
\def\JETP#1 {\Jl{Sov. Phys. JETP}{#1}}
\def\JHEP#1 {\Jl{JHEP}{#1}}
\def\JMP#1 {\Jl{J. Math. Phys.}{#1}}
\def\NPB#1 {\Jl{Nucl. Phys. B}{#1}}
\def\NP#1 {\Jl{Nucl. Phys.}{#1}}
\def\PLA#1 {\Jl{Phys. Lett. A}{#1}}
\def\PLB#1 {\Jl{Phys. Lett. B}{#1}}
\def\PRD#1 {\Jl{Phys. Rev. D}{#1}}
\def\PRL#1 {\Jl{Phys. Rev. Lett.}{#1}}

\def\lal{&&\nqq {}}

\def\beq{\begin{equation}}
\def\eeq{\end{equation}}
\def\bear{\begin{eqnarray}}
\def\bearr{\begin{eqnarray} \lal}
\def\ear{\end{eqnarray}}
\def\earn{\nonumber \end{eqnarray}}

\begin{document}
\thispagestyle{empty}
\twocolumn[

\vspace{1cm}

\Title{Scalarly charged particles and interparticle interaction with the Higgs potential}

\Author{Yu. G. Ignat'ev}
    {Institute of Physics, Kazan Federal University, Kremlyovskaya str., 16A, Kazan, 420008, Russia}

\Abstract
 {Asymptotically exact solutions are obtained for a spherically symmetric field with the Higgs potential generated by a point scalar charge, and a method for numerical integration of the equation for a scalar field with the Higgs potential of a point charge is proposed. Examples of numerical modeling of the equation of the scalar field of a single charge are given. With the help of the solution obtained, solutions of the relativistic equations of motion of a scalar charge in an external scalar field of the Higgs type of a singular scalar source are found and some unique properties of the interparticle scalar interaction are revealed.\\

 }

\bigskip

] 

\section*{Introduction}
Previously, the Author studied the question of interparticle interaction of scalarly charged particles with a quadratic interaction potential \cite{Ignat_12}.
The pur\-pose of this work is to study the interparticle inte\-rac\-tion of scalarly charged elementary particles in the Higgs potential model. This issue needs a detailed con\-si\-de\-ra\-tion for a deeper understanding of the mechanism of gra\-vi\-ta\-tional - scalar instability \cite{Ign_GC21_Un} -- \cite{YU_GC_4_22}, with the help of which it is possible to explain, in turn, the mechanism of formation of supermassive black holes in the early Universe.
The specificity of the task - the interaction of elementary scalar charged particles does not require taking into account gravitational interactions, like any other model of the interaction of elementary particles with masses below the Planck ones. However, it should be noted that in this case, the features of the interparticle interaction of scalarly charged particles at the macroscopic level significantly affect directly the dynamics of the gravitational field of matter, which was revealed in the above-cited works \cite{Ign_GC21_Un} -- \cite{YU_GC_4_22}. Note also that the essentially relativistic problem of the interparticle interaction of scalarly charged particles is solved below.
%
\section{Microscopic field equations with singular sources}
\subsection{Action functions for singular sources}
In this section, we briefly present the results of \cite{TMF_21} on the derivation of microscopic equations of motion of scalarly charged particles, adapting these results to the case of the absence of a gravitational field.

The action for a particle $(a)$ with a scalar charge $\mathrm{g}_a$ in a scalar field $\Phi$ can be written as an integral over the volume of the Riemannian space $R_4\equiv X$ \cite{TMF_21}:
\begin{equation}\label{S_a}
S_a =-\int\limits_X dX \int\limits_{\Gamma} D(x|x_a)\ m_{(a)}(s_a)ds_a,
\end{equation}
where
\begin{equation}\label{m_a}
m_{(a)}= \mathrm{g}_a\Phi
\end{equation}
-- particle dynamical mass, $D(x|x_a)$ -- invariant symmetric two-point $\delta$-function
{\small
\begin{equation}\label{D}
\!\!\!\!\int\limits_{X_2}D(x_1|x_2)F(x_2)dX_2=
\left\{\begin{array}{ll}
F(x_1), & x_1\in X_2;\\
0, & x_1 \not\in X_2,
\end{array}\right.
\end{equation}
}
$X_2\subset R_n$; $F(x)$ is an arbitrary tensor field on $R_n$, $dX=\sqrt{-g} dx^1\ldots dx^n$ is an invariant volume element of $R_n$, and $D(x_1|x_2 )=D(x_2|x_1)$. The internal integration in \eqref{S_a} is carried out over the own time $s_a$ along the entire trajectory of the particle $\Gamma$. This allows us to write down the total action integral for the system ``scalar field' + ``scalarly charged particles'' also as an integral over $X$:

\begin{eqnarray}\label{S3}
S=S_a+S_s=- \int\limits_X L_s dX \nonumber\\
-\int\limits_X dX\sum\limits_a \int\limits_{\Gamma_a}D(x|x_a)\ m_{(a)}(s_a)ds_a ,
\end{eqnarray}
where $L_s$ is the Lagrange function of the scalar field
\begin{eqnarray} \label{Ls}
L_s=\frac{1}{16\pi}(e \eta^{ik} \Phi_{,i} \Phi _{,k} -2V(\Phi),
\end{eqnarray}
\begin{eqnarray}
\label{Higgs}
V(\Phi)=-\frac{\alpha}{4} \left(\Phi^{2} -\frac{m^{2} }{\alpha}\right)^{2}
\end{eqnarray}
-- potential energy of the scalar field, $\alpha$ -- self-action constant, $m$ -- quantum mass, $e=\pm 1$ (sign ``+'' corresponds to the classical scalar field, sign ``-'' - - phantom).

\subsection{Euler - Lagrange equations}
Calculating the variation of the \eqref{S3} action with respect to the dynamic variables of the particles, $\delta_a S=\delta_aS_p$, and taking into account the dif\-fe\-ren\-tia\-tion property of the invariant $\delta$ -function (see \cite{TMF_21}), due to which the variation operation is also carried over to functions on the right $\delta D(x|y)F(y)=D(x|y)\delta F(y)$, as a result we get:
\[ \delta_a S=-\int\limits_X dX\sum\limits_a \int\limits_{\Gamma_a}D(x|x_a)\delta_a(m_{(a)}(s_a)ds_a).\]
Equating the variation $\delta_aS $ to zero, we obtain the Lagrangian equations of motion of scalarly charged particles
\begin{equation}\label{du_a/ds_m}
\frac{\delta u_a^i}{\delta s_a}=\partial_k\ln|m_{(a)}|\ \pi^{ik}(u_a),
\end{equation}
where $u_a^i=dx_a^i/ds_a$ is the velocity vector,
\begin{eqnarray}\label{pi^ik}
\pi^{ik}(u)=g^{ik}-u^iu^k\nonumber
\end{eqnarray}
is the projection tensor onto the $u^i$ direction and $\delta/\delta s_a$ is the absolute derivative (see, for example, \cite{Sing}).

Calculating now the variation of the action of particles in the scalar field $\delta S_p$, taking into account the definition of the dynamic mass of particles \eqref{m_a}, we similarly obtain:
\[\delta S_p=-\int\limits_X dX\sum\limits_a \mathrm{g}_{(a)}\int\limits_{\Gamma_a}D(x|x_a)_{(a)}\delta\Phi_r  ds_a. \]
Varying the actions of scalar fields over scalar fields, we find the total variation $\delta_\Phi S=\delta S_a+\delta_\Phi S_s$
\begin{eqnarray}
\delta_\Phi S= -\int\limits_X dX \delta\Phi \biggl(\sum\limits_a \mathrm{g}_{(a)} \int\limits_{\Gamma_a}D(x|x_a) ds_a\nonumber\\
+\frac{1}{8\pi}\bigl(e\Box\Phi+V'_{\Phi}\bigr) \biggr).\nonumber
\end{eqnarray}
Thus, we obtain microscopic equations of scalar fields:
\begin{equation}\label{Box(Phi)}
e\Box \Phi + V'_{\Phi} = -8\pi\sigma,
\end{equation}
Thus, we obtain microscopic equations of scalar fields:
\begin{equation}\label{sigma}
\sigma= \sum\limits_a \mathrm{g}_{(a)}\int D(x|x_a)ds_a
\end{equation}
is the microscopic density of the scalar charge of the system of particles.
\subsection{Energy - momentum tensors and conservation laws}
Calculating the variation of the action \eqref{S3} with respect to the metric tensor, we get:
microscopic energy-momentum tensor of scalarly charged particles, $T^i_{(p) k}$, --
\begin{equation}\label{T_p_m}
T^{ik}_{(p)}(x) = \sum\limits_a \int\limits_{\Gamma_a} m_{(a)}u^i_au^k_a(x|x_a)ds_a,
\end{equation}
and the energy-momentum tensor of scalar fields, $T^i_{(s)k}$, --
\begin{eqnarray}\label{T_s}
T^i_{(s)k}=\frac{1}{16\pi}\bigl( 2\Phi^{,i}\Phi_{,k} -\delta^i_k \Phi_{,j}\Phi^{,j} +2V(\Phi)\delta^i_k \bigr).
\end{eqnarray}

Calculating further the covariant divergences of these tensors, taking into account the symmetry of the invariant $\delta$ -function, we obtain:
\begin{eqnarray}\label{div_sing_T_p}
\nabla_kT^{ik}_{(p)}=\sum\limits_a \mathrm{g}_{(a)} \nabla^i\Phi\int\limits_{\Gamma_a} D(x|x_a) ds_a\Rightarrow\nonumber\\
\nabla_kT^{ik}_{(p)}=\sigma\nabla^i \Phi.
\end{eqnarray}

It is easy to see that due to \eqref{div_sing_T_p}, \eqref{Box(Phi)} and \eqref{T_s} the law of conservation of the total energy-momentum tensor of the system ``scalarly charged particles'' + ``scalar fields'' is automatically fulfilled '
\begin{equation}\label{nabla(Tp+Ts)}
\nabla_k T^{ik}=0 ; \qquad(T^{ik}=T^{ik}_{(p)}+T^{ik}_{(s)}),
\end{equation}
where the energy-momentum tensor of scalarly charged particles, $T^{ik}$, and the energy-momentum tensor of scalar fields, $T^{ik}_{(s)}$, are described
the formulas \eqref{T_p_m} and \eqref{T_s}, respectively.

\section{Scalar field of a point charge}
\subsection{Field equation for a single charge}

According to \eqref{Box(Phi)} and \eqref{sigma}, the scalar field with the Higgs potential $\Phi$ of a single point scalar charge $\mathrm{g}$ resting at
the origin of the spherical coordinate system is defined by the equation
\begin{equation}\label{spher_eq_e}
\frac{e}{r^2}\frac{d}{dr}\biggl(r^2\frac{d\Phi}{dr}\biggr)-m^2\Phi+\alpha\Phi^3=-8\pi \mathrm{g}\delta(\mathbf{r}).
\end{equation}
In this case, it is necessary to keep in mind the ratio:
\begin{equation}\label{1/r}
\frac{1}{r^2}\frac{d}{dr}\biggl(r^2\frac{d}{dr}\frac{1}{r}\biggr)=-4\pi\delta(\mathbf{r}).
\end{equation}

For $\alpha=m=0$, $e=+1$ the equation \eqref{spher_eq_e} has as its solution the Coulomb potential $\Phi=\mathrm{g}/r$, for $\alpha=0,m\not =0$ and $e=+1$ -- Yukawa potential
\begin{equation}\label{Yukava}
\Phi=\frac{2\mathrm{g}}{r}e^{-mr}.
\end{equation}
For $\alpha=0,m\not=0$ and $e=-1$ the solution of the corresponding equation is obtained from \eqref{Yukava} as its real part by replacing $m\to im$ and $g\to-\mathrm{ g}$:
\begin{equation}\label{Fantom_Yukava}
\Phi=-\frac{2\mathrm{g}}{r}\cos(mr).
\end{equation}
\subsection{Linear approximation}
Note that in vacuum the one-dimensional dynamical system with $\Phi=\Phi(x)$ corresponding to the equation \eqref{spher_eq_e} has two singular points (see, for example, \cite{YuKokh_TMF})
\begin{equation}\label{sing_points}
M_\pm:\; \Phi_\pm\equiv\pm\Phi_0=\pm\frac{m}{\sqrt{\alpha}},\; Z_\pm\equiv\Phi'_{\Phi_\pm}=0,
\end{equation}
which are attracting centers for a classical field ($e=+1$) and saddle (unstable) points for a phantom field ($e=-1$).
Thus, in the case of $e=+1$, according to the qualitative theory of dynamical systems, the solutions of the vacuum equation \eqref{spher_eq_e} as $r\to\infty$ must wrap around one of these singular points. In this case, all phase trajectories in the $\Phi,Z$ plane are divided into two symmetric flows, the trajectories of each of which wind around the corresponding singular point. Note, however, that the dynamical system corresponding to the equation \eqref{spher_eq_e} in spherical coordinates is not autonomous, therefore the results of the qualitative theory, strictly speaking, are not applicable to this system in sphe\-rical coordinates. On the other hand, any spherically symmetric field can be expanded in plane harmonics $\mathrm{e}^{i\mathbf{kr}}$ with a wave number, to each of which the conclusions of the qualitative theory of dynamical systems are quite applicable.
\Fig{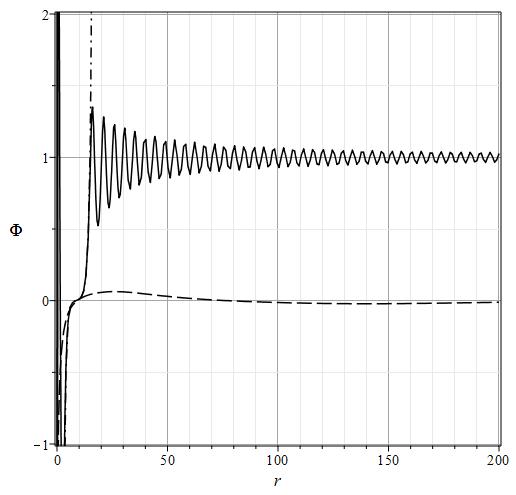}{8}{\label{ignatev-1}Scalar potential $\Phi(r)$ of the classical field ($e=+1$): dashed line -- $\alpha=0,m=1$, dashed line -- $\alpha=1,m=0 $; solid line -- $\alpha=1$, $m=1$.}
This fact is demonstrated by the graphs in Fig. \ref{ignatev-1} obtained as a result of numerical integration of the vacuum equation \eqref{spher_eq_e} with the same initial conditions $\Phi(10)=0.01$; $\Phi'_r(10)=0.01$ and pa\-ra\-meters $\alpha=0.1$ and $m=0.1$. Commenting on these graphs, we note, first, that although for $\alpha=0$ the vacuum equation \eqref{spher_eq_e} coincides with the equation for the Yukawa potential, under the chosen initial conditions its solution does not coincide with the Yukawa potential. Secondly, in the case of $\alpha\not=0$, the solutions of the \eqref{spher_eq_e} equation as $r\to\infty$ tend exactly to the corresponding stable singular points of the one-dimensional vacuum dynamical system \eqref{sing_points}.

Let us use this result and, choosing for definiteness the singular point $M_+$ at $\alpha\not\equiv0$, pass to a new field variable
$\varphi(r)$
\begin{equation}\label{F->f}
\Phi=\frac{m}{\sqrt{\alpha}}(1+\phi(r)),
\end{equation}
Then the equation \eqref{spher_eq_e} in the approximation linear in $\phi(r)\ll1$ takes the form:
\begin{equation}\label{spher_eq_phi_e}
\frac{e}{r^2}\frac{d}{dr}\biggl(r^2\frac{d\phi}{dr}\biggr)+2m^2\phi=-8\pi \frac{\mathrm{g\sqrt{\alpha}}}{m}\delta(\mathbf{r}).
\end{equation}

In the case of a phantom field ($e=-1$), this equation coincides with the equation for the classical Yukawa potential with substitutions $m\to \sqrt{2}m$, $g\to 2g\sqrt{\alpha}/m$. Making these substitutions in \eqref{Yukava} and \eqref{F->f}, we get for the phantom field
\begin{equation}\label{Phi_Lin_f}
\Phi\approx\displaystyle\frac{m}{\sqrt{\alpha}}\biggl(1+\frac{2g\sqrt{\alpha}}{mr}\mathrm{e}^{-\sqrt{2}mr}\biggr),\; (e=-1).
\end{equation}

In the case of a classical field, it is necessary to use the Yukawa phantom field potential \eqref{Fantom_Yukava} with appropriate substitutions, which gives
\begin{equation}\label{Phi_Lin_c}
\Phi\approx\frac{m}{\sqrt{\alpha}}\biggl(1+\frac{2g\sqrt{\alpha}}{mr}\cos\sqrt{2}mr\biggr),\; (e=+1).
\end{equation}

Note, however, that, as we noted above, the singular points \eqref{sing_points} of the corresponding one - dimensional vacuum dynamical system are points of stable equilibrium only for the classical scalar field ($e=+1$), while for the phantom field ($e= -1$) these points are saddle points. Therefore, in the case of a phantom field, the \eqref{Phi_Lin_f} solution is unstable and hardly has any physical meaning.

Note also that the singular points \eqref{sing_points} $\Phi=\pm m^2/\sqrt{\alpha}$ correspond to the zero value of the po\-ten\-tial energy \eqref{Higgs}: $V(\Phi_\pm)=0$ , i.e., correspond to the \emph{vacuum state of the system}.
\subsection{Numerical Integration}
Let us turn to the linear solution for the classical field \eqref{Phi_Lin_c}, which is stable, which can also be verified by direct numerical integration of the equation \eqref{spher_eq_e}. Asymptotic solution
\eqref{Phi_Lin_c} is also very useful for the numerical integration of the \eqref{spher_eq_e} singular equation, since there is no direct way to construct a numerical solution of this essentially non-linear equation with a singular source. Taking into account the fact that the solution \eqref{Phi_Lin_c}, firstly, is stable and asymptotically exact away from the singularity at $\phi(r)\to0$, we use this solution to set the initial conditions for the vacuum equation \eqref{spher_eq_e} away from source:
{\small
\begin{eqnarray}\label{ICS}
\!\!\!\Phi(R)=\frac{m}{\sqrt{\alpha}}\biggl(1+\frac{2g\sqrt{\alpha}}{mR}\cos\sqrt{2}mR\biggr),\; (mR\gg1);\nonumber\\
\!\!\!\Phi'(R)=-\frac{e}{R}\biggl(\sin\sqrt{2}mR +\frac{\cos\sqrt{2}mR}{\sqrt{2}mR}\biggr).
\end{eqnarray}
}
On Fig. \ref{ignatev-2} are shown in comparison of graphs of \eqref{Phi_Lin_c} asymptotic solution and numerical solution of \eqref{spher_eq_e} equation with initial conditions \eqref{ICS} at $R=10$. As can be seen from this figure, far from the source, the graphs of the potentials actually coincide, while near their behavior differs significantly.

\Fig{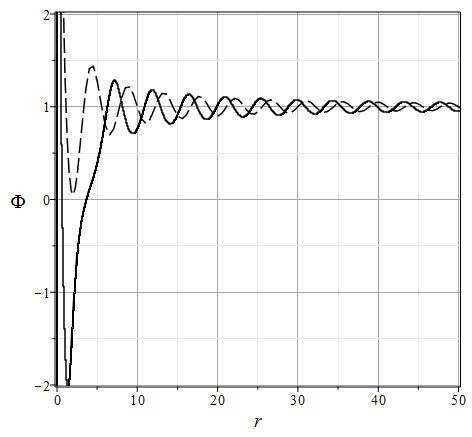}{8}{\label{ignatev-2}Scalar potential $\Phi(r)$ of the classical field ($e=+1$): the dashed line is the asymptotic solution \eqref{Phi_Lin_c}, the solid line is the numerical solution of the \eqref{Phi_Lin_c} equation with the initial conditions \eqref{ICS} for $\alpha=1,m=1$.}

Thus, we can state that a method has been obtained for the numerical solution of the field equation for the scalar Higgs field with a singular source.

\section{Interparticle scalar interactions}
Let us use the obtained solution to study the question of the interaction of two scalarly charged point particles. To do this, consider the motion of a test scalarly charged particle with a scalar charge $\mathrm{q}$ in a field of scalar charge $\mathrm{g}$, which we will describe by the asymptotic solution \eqref{Phi_Lin_c}. In this case, the metric will be assumed to be the Minkowski metric in the spherical coordinate system
\begin{equation}\label{Mink}
ds^2=dt^2-dr^2-r^2(d\theta^2+\sin^2\theta d\varphi^2).
\end{equation}
\subsection{Integrals of motion}
Let's turn to the equations of motion \eqref{du_a/ds_m}. Instead of directly solving these equations, let's turn to their integrals of motion, which have the form \cite{TMF_21}:
\begin{equation}\label{P_xi}
\mathbf{P}_\xi\equiv (\xi,P)=\mathrm{Const} \Leftarrow\!\!\Rightarrow \ \Lee{\xi}\Phi g_{ik}=0,
\end{equation}
where $\Lee{\xi}$ is the Lie derivative in the direction $\xi$ (see, for example, \cite{Petrov}). Here $P_i$ is a generalized momentum related to the velocity vector by the relation:
\begin{equation}\label{P_i}
P^i=\mathrm{q}\Phi u^i\equiv \mathrm{q}\Phi \frac{dx^i}{ds}.
\end{equation}
Under conditions of spherical symmetry, there are 4 Killing vectors $\xi^i$ \cite{Petrov}
\begin{eqnarray}\label{Killings}
\stackunder{1}{\xi}=(0,\sin\varphi,\mathrm{ctg}\,\theta\cos\varphi,0),\nonumber\\
\stackunder{2}{\xi}=(0,-\cos\varphi,\mathrm{ctg}\,\theta\sin\varphi,0),\nonumber\\
\stackunder{3}{\xi}=(0,0,-1,0);\; \stackunder{4}{\xi}=(0,0,0,1),
\end{eqnarray}
which correspond to 4 integrals of motion $\mathrm{P}_{(a)}=(P,\stackunder{a}{\xi})=\mathrm{Const}$ ($a=\overline{1,4}$ ). In addition, there is a mass integral:
\begin{equation}\label{P^2=m^2}
(P,P)=\mathrm{q}^2\Phi^2.
\end{equation}
These integrals are sufficient to completely describe the motion of a scalarly charged particle.

Thus, we get from \eqref{P_i} and \eqref{Killings}, taking into account \eqref{P^2=m^2}:
\begin{eqnarray}
\label{Int_1}
r^2(\sin\varphi\dot\theta+\sin\theta\cos\theta\cos\varphi \dot{\varphi})=\frac{\mathrm{P}_{(1)}}{\mathrm{E}_0};\\
\label{Int_2}
r^2(-\cos\varphi\dot\theta+\sin\theta\cos\theta\sin\varphi\dot{\varphi})=\frac{\mathrm{P}_{(2)}}{\mathrm{E}_0};\\
\label{Int_3}
r^2\sin^2\theta\dot\varphi=\frac{\mathrm{P}_{(3)}}{\mathcal{E}_0}\equiv\frac{\mathrm{L}_0}{\mathrm{E}_0};\\
\label{Int_4}
\mathrm{q}\Phi\frac{dt}{ds}=\mathrm{E}_0,\quad (\equiv\mathrm{P}_{(4)}),
\end{eqnarray}
where $\dot{f}\equiv df/dt$.
\subsection{Orbits}
Of the first three integrals \eqref{Int_1} -- \eqref{Int_3}, only two are functionally independent. The first two integrals \eqref{Int_1} -- \eqref{Int_2} ensure that the trajectory belongs to some plane passing through the origin (see, for example, \cite{Yu_Book}).
Choosing this plane as $\theta=\pi/2\Rightarrow \mathrm{P}_1=\mathrm{P}_2=0$, we obtain for \eqref{Int_3} the angular momentum conservation law $\mathrm{L}_0\equiv \mathrm{P}_{(3)}$:
\begin{equation}\label{L_0}
r^2\dot\varphi=\frac{\mathrm{L}_0}{\mathrm{E}_0}.
\end{equation}
Thus, from the mass integral \eqref{P^2=m^2} we get:
\begin{equation}\label{dr/dt}
\dot{r}^2=\frac{1}{\mathrm{E}^2_0}\biggl[\mathrm{E}^2_0-\biggl(g^2\Phi^2(r)+\frac{\mathrm{L}^2_0}{r^2}\biggr)\biggr],
\end{equation}
whence, due to the non-negativity of $\dot{r}^2$, we find the boundaries of the region of existence of motion:
\begin{equation}
\mathrm{E}_0^2-U^2(r)\equiv \mathrm{E}_0^2 -\biggl(\mathrm{g}^2\Phi^2(r)+\frac{\mathrm{L}^2_0}{r^2}\biggr)\geqslant0.
\end{equation}

From \eqref{dr/dt} we find the turning points of the trajectory (see, for example, \cite{Land_Mech}) at which $\dot{r}=0$:
\begin{equation}\label{dr/dt=0}
\mathrm{E}_0^2-U^2(r)\equiv \mathrm{E}_0^2 -\biggl(\mathrm{g}^2\Phi^2(r)+\frac{\mathrm{L}^2_0}{r^2}\biggr)=0.
\end{equation}

On Fig. \ref{ignatev-3} -- \ref{ignatev-4} graphs of the $\mathrm{E}_0^2-U^2(r)$ function versus energy and angular momentum are shown. As you can see, in the examples
depending on the energy and angular momentum, there can be zero, one, two, four, etc. turning points. With negative values of the ordinate, there is no movement.

\Fig{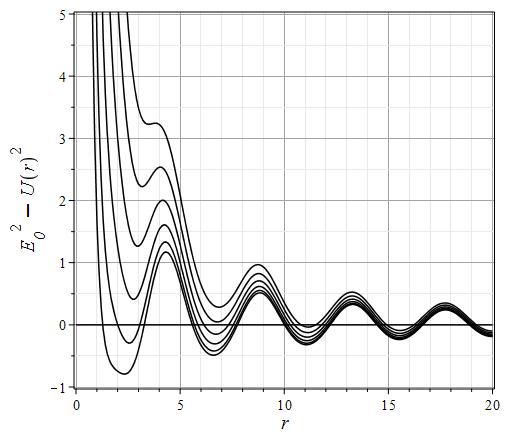}{7}{\label{ignatev-3}The left side of the equation \eqref{dr/dt=0} with respect to the asymptotic solution \eqref{Phi_Lin_c} for $\alpha=m=\mathrm{g}=\mathrm{q}=1$, $\mathrm{E}_0= 1$. From right to left: $\mathrm{L}_0=1$; $1$; $3$; $4$; $5$; $6$.}

\Fig{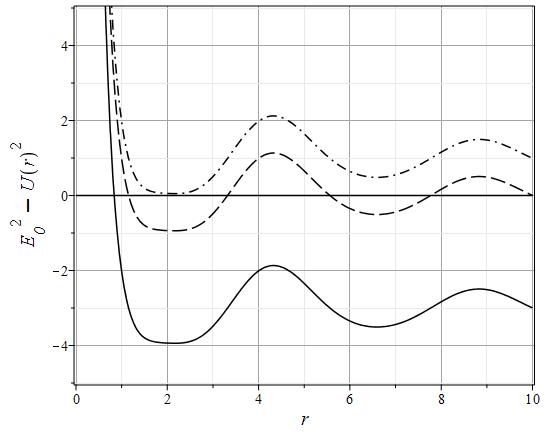}{7}{\label{ignatev-4}The left side of the equation \eqref{dr/dt=0} with respect to the asymptotic solution \eqref{Phi_Lin_c} for $\alpha=m=\mathrm{g}=\mathrm{q}=1$, $\mathrm{L}_0= 0.5$. From bottom to top: $\mathrm{E}_0=0.1$; $1$; $2$.}

\subsection{Circular orbits}
Circular orbits correspond to $r=r_0=\mathrm{Const}$. Then from \eqref{L_0} we immediately get:
\[\varphi=\frac{L_0}{r^2_0E_0}t.\]
From \eqref{dr/dt} we get for such orbits:
\begin{equation}\label{cicls}
g^2\Phi^2(r_0)+\frac{\mathrm{L}^2_0}{r_0^2}=\mathrm{E}^2_0
\end{equation}
-- the radiuses of the circular orbits are the same as the radii of the turning points (see Fig. \ref{ignatev-3} -- \ref{ignatev-4}). As we have seen, there can be at least several such points, as well as the corresponding solutions of the \eqref{cicls} equation. This is one of the non-standard features of the Higgs scalar interaction: several circular orbits can correspond to the same values of total energy and angular momentum.
\subsection{Numerical modeling}
Although it is convenient to carry out a general analysis of motion on the basis of first integrals of motion, for numerical modeling it is more convenient to use the second order equations of motion \eqref{du_a/ds_m}, since the first order differential equation \eqref{dr/dt} is rather inconvenient for numerical integration.\\ Passing in the equation \eqref{du_a/ds_m} to differentiation with respect to time $t$ using the formula \eqref{Int_4}, and also using the mass integral \eqref{P^2=m^2}, we reduce the equations of motion to the form of equations, allowed with respect to higher derivatives:
\begin{equation}
\ddot{r}=\frac{q^2\Phi^2\mathrm{L}_0^2}{\mathrm{E}_0^2r^3}+\frac{q^2\Phi\Phi'_r}{\mathrm{E}_0^2}\times\nonumber
\end{equation}
\begin{equation}\label{Sys_r}
\biggl[-3+\frac{q^2\Phi^2}{\mathrm{E}_0^2}+\frac{\mathrm{E}_0^2}{q^2\Phi^2}+\frac{\mathrm{L}_0^2}{r^2}\biggl(\frac{1}{\mathrm{E}_0^2}-\frac{1}{q^2\Phi^2}\biggr)\biggr];
\end{equation}
\begin{equation}
\label{Sys_varphi}
\dot{\varphi}=\frac{1}{r^2}.
\end{equation}

\eqref{Sys_r} -- \eqref{Sys_varphi} equations must be solved with initial conditions
\begin{eqnarray}
r(0)=r_0,\qquad \varphi(0)=0;\nonumber\\
\left.\frac{dr}{dt}\right|_{t=0}=\pm \sqrt{1-\frac{1}{\mathrm{E}^2_0}\biggl(g^2\Phi^2(r_0)+\frac{\mathrm{L}^2_0}{r_0^2}\biggr)},
\end{eqnarray}
where the sign ``$+$"" cor\-res\-ponds to the initial speed directed away from the center, the sign ``$-$'' cor\-res\-ponds to the initial speed directed towards the center.

On Fig. \ref{ignatev-5} shows the results of numerical moderling of the trajectory of a scalar charged particle with parameters $\mathrm{q}=1,\mathrm{L}_0=1,\mathrm{E}_0=1$ in the field of a scalar charge with parameters $\alpha=1,\mathrm{g}=1,m=1$,
\Fig{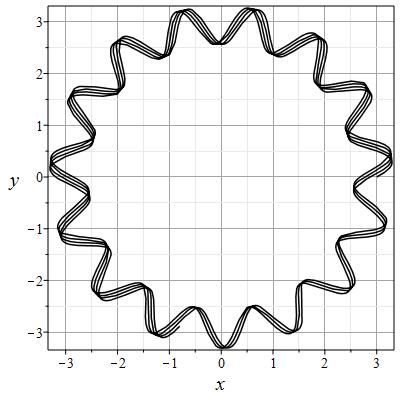}{7}{\label{ignatev-5}Trajectory of motion of a scalarly charged particle with charge $\mathrm{q}=1$ in the field of asymptotic solution \eqref{Phi_Lin_c} for $\alpha=m=\mathrm{g}=\mathrm{q}=1$, $\mathrm {L}_0=1,\mathrm{E}_0=1$.}
From this figure, firstly, it can be seen that the trajectory is enclosed between two turning points $r_\pm$ (or rather, circles), and secondly, the trajectory is not closed. As is known, in the classical problem of motion in a central field, the trajectory can be closed only in the case of the potential $U(r)\backsim r^{-1},\ r^{-2}$ (see, for example, \cite{Land_Mech }).
\subsection{Radial motion}
Radial motion corresponds to zero angular momentum $\mathrm{L}_0=0$. From \eqref{L_0} we get $\varphi=\varphi_0$. On Fig. \ref{ignatev-6} graphs of the trajectories of a particle initially falling on a scalar charge are shown
($\dot{r}|_{t=0}<0$).

As can be seen from this figure \ref{ignatev-7}, in the case of sufficiently high energies of the incident particle $\mathrm{E}_0>1.25$, the field of the central charge repels the incident particle. However, at $\mathrm{E}_0=1.25$ this type of motion changes to periodic oscillations of the particle along the radius. On Fig. \ref{ignatev-7} shows a similar example of the oscillatory motion of a test particle along the radius connecting it with the central charge. It should be noted that this unique property is inherent in the scalar Higgs interaction. This property shows that at a small total energy, repulsive forces act near the central charge, which are replaced by attraction when moving away from the charge.

\Fig{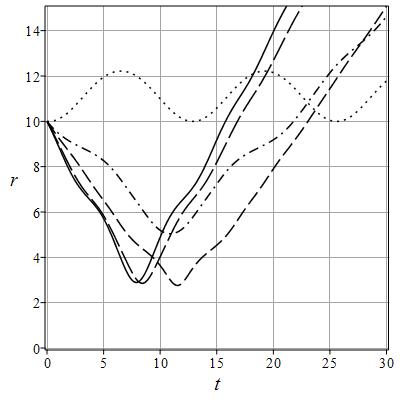}{7}{\label{ignatev-6}Radial motion of a scalarly charged particle with charge $\mathrm{q}=1$ in the field of asymptotic solution \eqref{Phi_Lin_c} for $\alpha=m=\mathrm{g}=\mathrm{q}=1$, $\mathrm {L}_0=0$, $r_0=10$: solid line -- $\mathrm{E}_0=5$; long-dashed -- $2.5$; dashed -- $1.5$; dash - dotted - $1.25$; dashed -- $1$.}
\Fig{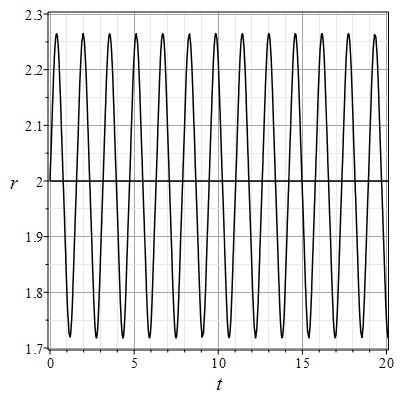}{7}{\label{ignatev-7}Radial motion of a scalarly charged particle with charge $\mathrm{q}=1$ in the field of \eqref{Phi_Lin_c} asymptotic solution for $\alpha=m=\mathrm{g}=\mathrm{q}=1$, $\mathrm {L}_0=0$, $r_0=2$; $\mathrm{E}_0=0.1$.}
\subsection*{Conclusion}
Summing up the work, we summarize its main results.\\[8pt]
\noindent 1. An asymptotically exact solution of the scalar field equation with the Higgs potential for a single singular scalar charge is obtained.\\[8pt]
\noindent 2. Using this solution, a method for numerically solving the scalar field equation with the Higgs potential for a single singular scalar charge is constructed.\\[8pt]
\noindent 3. The integrals of motion of a test scalar charge in a field with the Higgs potential of a single singular charge are obtained.\\[8pt]
\noindent 4. With the help of the found integrals of motion, the admissible regions of motion of a test scalar charge are investigated and turning points are found, which turn out to be several.\\[8pt]
\noindent 5. It is shown that several circular orbits determined by the initial conditions can correspond to the same values of the angular momentum and total energy.\\[8pt]
\noindent 6. Numerical models of the motion of a scalarly charged particle in a field with the Higgs potential of a single singular charge are constructed. It is shown that the trajectories are open rosettes.\\[8pt]
\noindent 7. The radial incidence of a scalarly charged particle on a single singular charge is studied. It is shown that at high values of the total energy the test particle is repelled by a scalar charge, and at low values it performs radial oscillatory motions.

Thus, some unique properties of the interparticle scalar interaction have been established in the work. In particular, the existence of oscillatory radial motion indicates that repulsion prevails at small distances, while attraction prevails at large distances.

\section*{Founding}
The work was carried out at the expense of a subsidy allocated as part of the state support of the Kazan (Volga Region) Federal University in order to increase its competitiveness among the world's leading scientific and educational centers.


\begin{thebibliography}{15}
%
\bibitem{Ignat_12}
Yu. G. Ignatyev (Ignat'ev), \emph{Russ. Phys. J.}, \textbf{55}, 1345 (2013); arXiv:1307.2509[gr-qc].
%
\bibitem{Ign_GC21_Un}
Yu. G. Ignat'ev, \emph{Gravit. Cosmol}., \textbf{28}, 25 (2022); arXiv:2203.11948 [gr-qc].
%
%
\bibitem{Yu_GC_3_22}
Yu. G. Ignat'ev, \emph{Gravit. Cosmol}., \textbf{28}, 275 (2022); arXiv:2207.05066 [gr-qc].
%
\bibitem{YU_GC_4_22}
Yu. G. Ignat'ev, \emph{Gravit. Cosmol}., \textbf{28}, 375 (2022); arXiv:2211.14507v1 [gr-qc].
%
%
\bibitem{TMF_21}
Yu. G. Ignat'ev and D. Yu. Ignat'ev,  \emph{Theoret. and Math. Phys.}, \textbf{209},  1437 (2021); arXiv:2111.00492 [gr-qc].
%
\bibitem{Sing}
J. L. Sing, ``Relativity: The General Theory'' (North-Holland Publ. Comp., Amsterdam, 1960).
%
%
\bibitem{YuKokh_TMF}
Yu. G. Ignat'ev and I. A. Kokh, \emph{Theor. Math. Phys.}, \textbf{207:1}, 514 (2021); arXiv:2104.01054 [gr-qc].
%
\bibitem{Petrov}
A. Z. Petrov, ``New Methods in General Relativity Theory'',  (Publ. Nauka, Moskow, 1966).
%
\bibitem{Yu_Book}
Yu. G. Ignat'ev, A. A. Agafonov, ``Mathematical Models of Theoretical Physics'', (Publ. House of Kazan University, Kazan, 2017).
%
\bibitem{Land_Mech}
L. D. Landau and E. M. Lifshitz, ``The Theoretical Physics, Vol. 1. Mechanics'', (Publ. Nauka, Moskow, 1988).
%
%
\end{thebibliography}
\end{document}